\documentclass[aps,prl,showpacs,twocolumn,groupedaddress,superscriptaddress]
{revtex4}
\usepackage{graphicx}
\usepackage{dcolumn} 
\usepackage{longtable} 
\usepackage{amsmath} 

\begin{document} 

\title{{\it Ab initio} calculations of charge symmetry breaking in the $A=4$ 
hypernuclei} 

\author{Daniel Gazda}\thanks{gazda@ujf.cas.cz}
\affiliation{Nuclear Physics Institute, 25068 \v{R}e\v{z}, Czech Republic}
\affiliation{ECT$^{\ast}$, Villa Tambosi, 38123 Villazzano (Trento), Italy}
\affiliation{Department of Fundamental Physics, Chalmers University of 
Technology, SE-412 96 G\"{o}teborg, Sweden}\author{Avraham Gal}
\thanks{avragal@savion.huji.ac.il}\affiliation{Racah Institute of Physics, 
The Hebrew University, Jerusalem 91904, Israel}

\date{\today}

\begin{abstract}

We report on {\it ab initio} no-core shell model calculations of the mirror 
$\Lambda$ hypernuclei $_{\Lambda}^4$H and $_{\Lambda}^4$He, using the 
Bonn-J\"{u}lich leading-order chiral effective field theory hyperon-nucleon 
potentials plus a charge symmetry breaking $\Lambda$-$\Sigma^0$ mixing vertex. 
In addition to reproducing rather well the $0^{+}_{\rm g.s.}$ and $1^{+}_{
\rm exc.}$ binding energies, these four-body calculations demonstrate for the 
first time that the observed charge symmetry breaking splitting of mirror 
levels, reaching hundreds of keV for $0^{+}_{\rm g.s.}$, can be reproduced 
using realistic theoretical interaction models, although with a non-negligible 
momentum cutoff dependence. Our results are discussed in relation to recent 
measurements of the $_{\Lambda}^4$H($0^{+}_{\rm g.s.}$) binding energy 
[MAMI A1 Collaboration, Phys. Rev. Lett. {\bf 114}, 232501 (2015)] and the 
$_{\Lambda}^4{\rm He}(1^{+}_{\rm exc.})$ excitation energy [J-PARC E13 
Collaboration, Phys. Rev. Lett. {\bf 115}, 222501 (2015)]. 

\end{abstract}

\maketitle

\begin{figure}[htbp] 
\begin{center} 
\includegraphics[width=4.8cm]{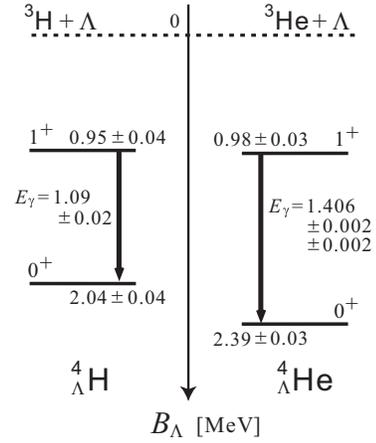} 
\caption{(${_{\Lambda}^4{\rm H}},{_{\Lambda}^4{\rm He}}$) mirror hypernuclei 
level diagram. The $0^+_{\rm g.s.}$ $\Lambda$ separation energies 
$B_{\Lambda}$, loosely termed $\Lambda$ binding energies, are from emulsion 
work \cite{davis05}, and the $1^{+}_{\rm exc.}$ $B_{\Lambda}$ values follow 
from $\gamma$-ray measurements of the excitation energies $E_{\gamma}$ 
\cite{E13}.} 
\label{fig:exp} 
\end{center} 
\end{figure}

\noindent 
{\bf Introduction.}~~Charge symmetry in hadronic physics is broken in QCD 
by the up-down light quark mass difference and by the up and down quark 
QED interactions. Recent lattice QCD+QED simulations of octet baryon mass 
differences within isospin multiplets, such as the neutron-proton mass 
difference $\Delta_{np}$ which vanishes in the limit of charge symmetry, 
account nicely for the observed charge symmetry breaking (CSB) pattern in 
the lowest-mass nonstrange as well as strange baryon spectrum \cite{LQCD15}. 
A comparable level of precision in reproducing theoretically CSB 
effects in the baryon-baryon interaction is lacking \cite{LQCD13}. 
In practice, introducing two charge-dependent contact interaction terms 
in chiral effective field theory (EFT) applications, one is able at 
next-to-next-to-next-to-leading order (N3LO) to account quantitatively 
for the charge dependence of the low-energy nucleon-nucleon ($NN$) 
scattering parameters~\cite{entem03}. For strangeness $S=-1$, however, 
given that low-energy $\Lambda p$ cross sections are poorly known and 
$\Lambda n$ scattering data do not exist, the available chiral EFT 
hyperon-nucleon ($YN$) interactions \cite{polinder06,haidenbauer13} 
do not include charge-dependent interaction terms. Potentially unique 
information on CSB in the $\Lambda N$ interaction and in $\Lambda$ 
hypernuclei is provided by the large $\Lambda$ separation-energy difference 
$\Delta B^{J=0}_{\Lambda}$=350$\pm$60~keV~\cite{davis05} in the $A=4$ 
mirror hypernuclei $0^+$ ground states (g.s.) and the apparently negligible 
difference $\Delta B^{J=1}_{\Lambda}$ in the $1^+$ excited states \cite{E13}, 
see Fig.~\ref{fig:exp}. Here, $\Delta B^{J}_{\Lambda}\equiv B^{J}_{\Lambda}
(_{\Lambda}^4{\rm He})-B^{J}_{\Lambda}(_{\Lambda}^4{\rm H})$. The recent 
precise measurement of the $_{\Lambda}^4{\rm H}_{\rm g.s.}\to{^4{\rm He}}+
\pi^-$ decay at the Mainz Microtron (MAMI) \cite{MAMI15} reaffirms 
a substantial CSB g.s. splitting $\Delta B^{J=0}_{\Lambda}$=270$\pm$95~keV, 
which is consistent with the emulsion value cited above. Note that 
$\Delta B^{J=0}_{\Lambda}$ is considerably larger than the $\approx$70~keV 
assigned to CSB splitting in the mirror core nuclei $^3$H and 
$^3$He \cite{miller06}. 

\begin{figure}[t!] 
\begin{center} 
\includegraphics[width=60mm]{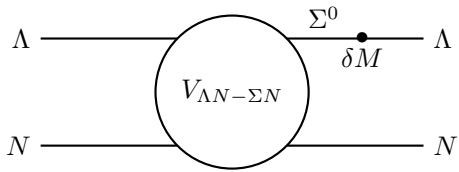} 
\caption{CSB $\Lambda N$ interaction diagram describing a SI $V_{\Lambda N - 
\Sigma N}$ interaction followed by a CSB $\Lambda-\Sigma^0$ mass-mixing 
vertex.} 
\label{fig:diag_csbls_h} 
\end{center} 
\end{figure}

Dalitz and von Hippel \cite{DvH64} suggested that the SU(3) octet 
$\Lambda_{I=0}$ and $\Sigma^0_{I=1}$ hyperons are admixed in the physical 
$\Lambda$ hyperon, thereby generating a CSB direct $\Lambda N$ potential 
$V_{\rm CSB}$ that consists of isovector meson exchanges, notably a long-range 
one-pion exchange (OPE) component. Although these exchanges are forbidden in 
the $\Lambda N$ channel by the strong interactions (SI), they do contribute 
strongly to the $\Lambda N\leftrightarrow\Sigma N$ coupling potential. 
Quite generally, the matrix element of $V_{\rm CSB}$ arising from 
$\Lambda-\Sigma^0$ mixing is related to the SI $I_{NY}=1/2$ matrix 
element $\langle N\Sigma|V_{\rm SI}|N\Lambda\rangle$ by \cite{gal15} 
\begin{equation} 
\langle N\Lambda|V_{\rm CSB}|N\Lambda\rangle = -0.0297\,\tau_{Nz}\,
\frac{1}{\sqrt{3}}\,\langle N\Sigma|V_{\rm SI}|N\Lambda\rangle , 
\label{eq:OME} 
\end{equation} 
where 
$\tau_{Nz}=\pm 1$ for protons and neutrons, respectively, 
and the space-spin structure of this $N\Sigma$ state is taken identical 
with that of the $N\Lambda$ state embracing $V_{\rm CSB}$. The CSB scale 
coefficient 0.0297 in (\ref{eq:OME}) follows from the $\Lambda-\Sigma^0$ 
mass-mixing matrix element \cite{gal15a} 
\begin{equation} 
\langle\Sigma^0|\delta M|\Lambda\rangle=\frac{1}{\sqrt 3}
(\Delta_{\Sigma^0\Sigma^+}-\Delta_{np})=1.14(5)~{\rm MeV}   
\label{eq:deltaM} 
\end{equation} 
and has been used in all previous CSB works listed below. A visualization 
of Eq.~(\ref{eq:OME}) is provided by the CSB $\Lambda N$ interaction diagram 
of Fig.~\ref{fig:diag_csbls_h}, where the $V_{\Lambda N - \Sigma N}$ blob 
represents {\it any} SI isovector meson exchange or contact term such as 
introduced in chiral EFT models \cite{polinder06}.  

Precise four-body calculations using the Nijmegen soft-core realistic meson 
exchange $YN$ interaction models NSC97$_{\rm e,f}$~\cite{NSC97}, which include 
charge-dependent interactions induced by $\Lambda-\Sigma^0$ mixing and meson 
mixings, produced at most 30\% of the observed CSB g.s. splitting $\Delta 
B^{J=0}_{\Lambda}$ \cite{nogga01,nogga02,haidenbauer07,nogga13,nogga14}. 
Below we comment on this insufficiency. More recent Nijmegen \cite{ESC10} or 
quark-cluster \cite{kyoto07} models have not been used in four-body studies. 
With SI $\Lambda N\leftrightarrow\Sigma N$ potential energy contributions 
of order 10~MeV \cite{nogga01}, and with a CSB scale of order 3\%, 
Eq.~(\ref{eq:OME}) could yield CSB contributions of order 300~keV. 
Reproducing the observed CSB splitting poses a challenge for microscopic 
$YN$ interaction models. 

In this Letter we report on detailed {\it ab initio} no-core shell model 
(NCSM) calculations of the $A=4$ $\Lambda$ hypernuclei that employ the SI 
Bonn-J\"{u}lich LO chiral EFT $YN$ interaction potentials \cite{polinder06}, 
plus a CSB $\Lambda-\Sigma^0$ mixing interaction potential $V_{\rm CSB}$ 
generated by applying Eq.~(\ref{eq:OME}) to each one of the $\Lambda N
\leftrightarrow\Sigma N$ $V_{\rm SI}$ components in this LO version. CSB meson 
mixings, with negligible contributions in the $A$=4 hypernuclei \cite{coon99}, 
are disregarded here. In addition to reproducing reasonably well the 
$0^{+}_{\rm g.s.}$ and $1^{+}_{\rm exc.}$ binding energies, these four-body 
calculations establish for the first time as large CSB splittings $\Delta 
B^{J=0}_{\Lambda}$ as suggested by experiment, see Fig.~\ref{fig:exp}, 
although with a non-negligible cutoff dependence. We also discuss possible 
implications to the recent Bonn-J\"{u}lich-Munich NLO chiral EFT $YN$ 
interaction model \cite{haidenbauer13}. 

\noindent 
{\bf Methodology.}~~The nuclear NCSM technique used in the present four-body 
calculations employs realistic two-body and three-body model interactions 
and is formulated in a translationally invariant Jacobi-coordinate 
harmonic-oscillator (HO) basis \cite{navratil00}. Antisymmetrization with 
respect to nucleons is exercised in order to satisfy the Pauli principle. The 
resulting Hamiltonian is diagonalized in finite four-body HO bases, admitting 
all HO excitation energies $N\hbar\omega$, $N\leq N_{\rm max}$, up to 
$N_{\rm max}$ HO quanta. Extrapolated energy values $E(\omega)$, $N_{\rm max}
\to\infty$, are obtained by fitting an exponential function to $E(N_{\rm max},
\omega \text{~fixed})$ sequences in the vicinity of the variational minima 
with respect to the HO basis frequency $\omega$. The reliability of such 
extrapolations is then reflected in the independence of $E(\omega)$ of the 
frequency $\omega$. 

\begin{table}[t!]
\caption{Cutoff dependence of $\Lambda$ separation energies $B^{J}_{\Lambda}$  
in $_{\Lambda}^4{\rm H}$ and $_{\Lambda}^4{\rm He}$ (all in MeV) from {\it ab 
initio} NCSM calculations at $\hbar\omega$=30(32)~MeV for $J$=0(1), using N3LO 
(LO) chiral $NN$ ($YN$) interactions \cite{entem03} (\cite{polinder06}) plus 
Coulomb interactions, and $V_{\rm CSB}$ generated by Eq.~(\ref{eq:OME}) from 
the LO SI $YN$ potentials. Experimental values are from Fig.~\ref{fig:exp}.}
\begin{ruledtabular} 
\begin{tabular}{lccccc} 
Cutoff & 550 & 600 & 650 & 700 & Experiment \\ 
\hline 
$B^{J=0}_{\Lambda}(_{\Lambda}^4{\rm H})$ & 2.556 & 2.308 & 2.154 & 2.196 & 
2.04$\pm$0.04 \\ 
$B^{J=0}_{\Lambda}(_{\Lambda}^4{\rm He})$ & 2.586 & 2.444 & 2.398 & 2.490 & 
2.39$\pm$0.03 \\ 
$B^{J=1}_{\Lambda}(_{\Lambda}^4{\rm H})$ & 1.744 & 1.359 & 1.067 & 0.877 & 
0.95$\pm$0.04 \\ 
$B^{J=1}_{\Lambda}(_{\Lambda}^4{\rm He})$ & 1.572 & 1.166 & 0.839 & 0.654 & 
0.98$\pm$0.03 \\ 
\end{tabular} 
\end{ruledtabular} 
\label{tab:B_L(A=4)} 
\end{table} 

This NCSM technique, extended recently to light hypernuclei 
\cite{gazda14,wirth14}, is applied here to the $A$=4 mirror hypernuclei 
using chiral N3LO $NN$ and N2LO $NNN$ interactions \cite{entem03,navratil07}, 
respectively, both with a momentum cutoff of 500 MeV. These together with 
the Coulomb interaction reproduce the binding energies of the $A$=3 core 
nuclei. For the SI $YN$ coupled-channel potentials $V_{\rm SI}$, we use the 
Bonn-J\"{u}lich LO chiral EFT SU(3)-based model with cutoff momenta $\Lambda$ 
from 550 to 700 MeV \cite{polinder06} plus $V_{\rm CSB}$ evaluated from 
$V_{\rm SI}$ by using Eq.~(\ref{eq:OME}). Baryon mass differences within 
isomultiplets are incorporated. The reported calculations consist of 
fully converged $^3$H and $^3$He binding energies, and ($_{\Lambda}^4$H, 
$_{\Lambda}^4$He) $0^+_{\rm g.s.}$ and $1^{+}_{\rm exc.}$ binding energies 
extrapolated to infinite model spaces from $N_{\rm max}=18(14)$ for $J=0(1)$. 
The $NNN$ interaction is excluded from the calculations reported here, 
in order to save computing time, after verifying that its inclusion makes 
a difference of only a few keV in the calculation of the CSB splittings 
$\Delta B^{J}_{\Lambda}$ for both $J=0,1$. 
\newline 
\noindent
{\bf Results.}~~The cutoff dependence of $\Lambda$ separation energies in both 
$A$=4 mirror hypernuclei, obtained from NCSM calculations with LO chiral EFT 
coupled-channel $YN$ potentials \cite{polinder06} and $V_{\rm CSB}$ from 
Eq.~(\ref{eq:OME}), is shown in Table~\ref{tab:B_L(A=4)}. We used 
$N_{\rm max}\to\infty$ extrapolated binding-energy values for the 
$_{\Lambda}^4{\rm He}$ and $_{\Lambda}^4{\rm H}$ $J$=0(1) levels at fixed 
$\hbar\omega$=30(32)~MeV, which is where the absolute variational minima 
occur for $\Lambda$=550 and 600~MeV. For higher values of $\Lambda$ the 
four-body absolute variational minima occur at slightly higher $\hbar\omega$ 
values. Although the spread of $B^{J}_{\Lambda}(\hbar\omega)$ values for 
a given cutoff momentum is of the order of 100~keV, it is considerably smaller 
and in fact marginal for the CSB splittings $\Delta B^{J}_{\Lambda}$ on which 
we focus here, as demonstrated by Fig.~4 below. 

The $\Lambda$ separation energies listed in Table~\ref{tab:B_L(A=4)} 
show a moderate cutoff dependence for the $0^+_{\rm g.s.}$ mirror levels 
and a stronger dependence for the $1^{+}_{\rm exc.}$ mirror levels, 
with mean values for their charge-symmetric (CS) averages given by 
${\overline B}^{\rm CS}_{\Lambda}(0^+_{\rm g.s.})$=2.39$^{+0.18}_{-0.12}$~MeV 
and ${\overline B}^{\rm CS}_{\Lambda}(1^{+}_{\rm exc.})$=1.16$^{+0.50}_{-0.39}
$~MeV which compare well with the CS-averaged experimental values derived from 
the last column in Table~\ref{tab:B_L(A=4)}. Furthermore, considering NCSM 
$N_{\rm max}\to\infty$ extrapolation uncertainties, our CS-averaged 
$B_{\Lambda}$ values are in fair agreement with those reported 
in other four-body calculations using CS LO $YN$ chiral EFT 
interactions \cite{haidenbauer07,nogga13,nogga14,gazda14,wirth14}. 
A detailed analysis of calculational uncertainties will be given elsewhere. 

\begin{figure}[t!] 
\begin{center} 
\includegraphics[width=80mm]{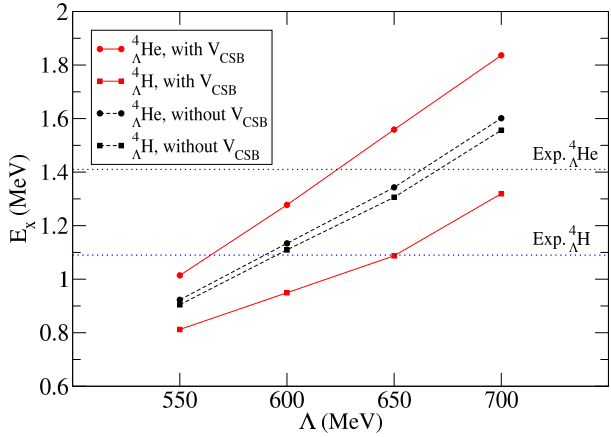}
\caption{(color online). Cutoff momentum dependence of excitation energies 
$E_{\rm x}$(0$^+_{\rm g.s.}$$\to$1$^+_{\rm exc.}$) in $_{\Lambda}^4{\rm H}$ 
(squares) and $_{\Lambda}^4{\rm He}$ (circles) in {\it ab initio} NCSM 
calculations, at $\hbar\omega$=30(32)~MeV for $J$=0(1), for LO chiral EFT 
coupled-channel $YN$ potentials \cite{polinder06} with (solid lines) and 
without (dashed lines) $V_{\rm CSB}$ derived from these SI potentials using 
Eq.~(\ref{eq:OME}). The dotted horizontal lines denote $E_{\rm x}$ values 
from $\gamma$-ray measurements \cite{E13}.} 
\label{fig:ex-l_d} 
\end{center} 
\end{figure} 

Shown in Fig.~\ref{fig:ex-l_d} by solid lines is the cutoff momentum 
dependence of the $0^+_{\rm g.s.}\to 1^{+}_{\rm exc.}$ excitation 
energies $E_{\rm x}$ formed from the $B_{\Lambda}$ values listed 
in Table~\ref{tab:B_L(A=4)} for both $A$=4 mirror hypernuclei. 
As observed in several few-body calculations of $s$-shell 
hypernuclei \cite{GL88,akaishi00,hiyama02,akaishi02}, $E_{\rm x}$ is 
strongly correlated with the $\Lambda N\leftrightarrow\Sigma N$ coupling 
potential which in the present context, through $\Lambda-\Sigma^0$ mixing, 
gives rise to CSB splittings of the $A$=4 mirror levels. 

Figure \ref{fig:ex-l_d} demonstrates a steady rise of both 
$E_{\rm x}(_{\Lambda}^4{\rm He})$ and $E_{\rm x}(_{\Lambda}^4{\rm H})$ 
as a function of the cutoff momentum $\Lambda$, with a CS-averaged value 
${\overline E}_{\rm x}^{\rm CS}$=1.23$^{+0.35}_{-0.32}$~MeV compared to 
1.25$\pm$0.02~MeV deduced from the two $\gamma$-ray energies shown in 
Fig.~\ref{fig:exp}. A steady rise is also observed in the difference 
$\Delta E_{\rm x}^{\rm CSB}$ with a mean value 380$^{+140}_{-180}$~keV 
compared to 320$\pm$20~keV, again from Fig.~\ref{fig:exp}. In agreement with 
previous calculations \cite{nogga01,nogga02,haidenbauer07,nogga13,nogga14}, 
residual CSB contributions of up to 30~keV from electromagnetic mass 
differences, mostly of $\Sigma$ hyperons, and from the increased Coulomb 
repulsion in the $^3$He core of $_{\Lambda}^4{\rm He}$, survive upon 
switching off $V_{\rm CSB}$, as demonstrated by the slight difference 
between the two middle dashed lines in the figure. 

\begin{figure}[t!] 
\begin{center} 
\includegraphics[width=80mm]{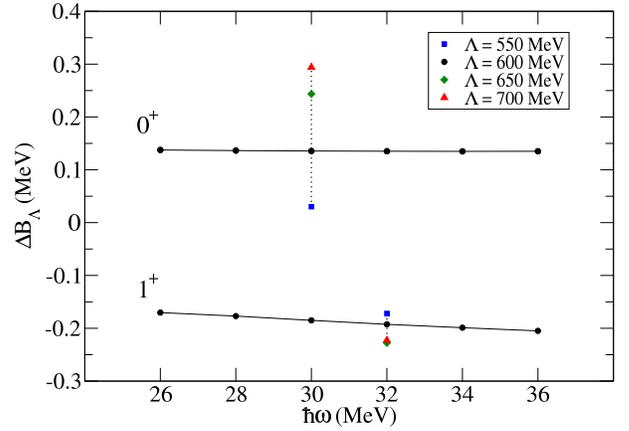} 
\caption{(color online). Dependence of the separation-energy differences 
$\Delta B_{\Lambda}$ between $_{\Lambda}^4{\rm He}$ and $_{\Lambda}^4{\rm H}$, 
for $0^+_{\rm g.s.}$ (upper curve) and for $1^+_{\rm exc.}$ (lower curve) 
on the HO $\hbar\omega$ in {\it ab initio} NCSM calculations using LO chiral 
EFT coupled-channel $YN$ potentials with cutoff momentum $\Lambda$=600~MeV 
\cite{polinder06} plus $V_{\rm CSB}$ derived from these SI potentials using 
Eq.~(\ref{eq:OME}). Results for other values of $\Lambda$ are shown at 
$\hbar\omega$=30(32)~MeV for $J$=0(1).}  
\label{fig:csb-om} 
\end{center} 
\end{figure}  

In Fig.~\ref{fig:csb-om} we show the $\hbar\omega$ dependence of 
separation-energy differences $\Delta B^{J}_{\Lambda}$ between $_{\Lambda}^4{
\rm He}$ and $_{\Lambda}^4{\rm H}$ levels of a given spin $J$, for $0^+_{
\rm g.s.}$ and $1^{+}_{\rm exc.}$, using $N_{\rm max}\to\infty$ extrapolated 
values for the four possible binding energies which are calculated for a 
cutoff $\Lambda$=600~MeV and including $V_{\rm CSB}$ from Eq.~(\ref{eq:OME}). 
Extrapolation uncertainties for $\Delta B^{J}_{\Lambda}$ are about 20~keV. 
The variation of $\Delta B^{J=0}_{\Lambda}$ in the spanned $\hbar\omega$ 
range amounts to a few keV, whereas that of $\Delta B^{J=1}_{\Lambda}$ is 
larger, amounting to $\sim$30~keV. It is worth noting that the difference 
$\Delta B^{J=0}_{\Lambda}-\Delta B^{J=1}_{\Lambda}$ between the upper and 
lower curves assumes at $\Lambda$=600~MeV the value 0.33$\pm$0.04~MeV, 
in perfect agreement with the difference $E_{\gamma}(_{\Lambda}^{4}{\rm He})
-E_{\gamma}(_{\Lambda}^{4}{\rm H})=0.32\pm 0.02$~MeV between the two $\gamma$ 
ray energies shown in Fig.~\ref{fig:exp}. The figure also demonstrates 
a strong cutoff dependence of $\Delta B^{J=0}_{\Lambda}$, varying between 
30 and 300~keV upon increasing $\Lambda$, together with a considerably weaker 
cutoff dependence of $\Delta B^{J=1}_{\Lambda}$, varying between $-$170 
and $-$230~keV. Note that $\Delta B^{J=0}_{\Lambda}$ comes out invariably 
positive, whereas $\Delta B^{J=1}_{\Lambda}$ is robustly negative. With mean 
values ${\overline{\Delta B}}^{J=0}_{\Lambda}$=176$^{+118}_{-146}$~keV and 
${\overline{\Delta B}}^{J=1}_{\Lambda}=-204^{+32}_{-24}$~keV, the mean 
values ${\overline{\Delta B}}^J_{\Lambda}$ satisfy 
\begin{equation} 
{\overline{\Delta B}}^{J=1}_{\Lambda} \approx 
-\,{\overline{\Delta B}}^{J=0}_{\Lambda} < 0. 
\label{eq:minus} 
\end{equation} 
\noindent
{\bf Discussion.}~~To understand the CSB pattern Eq.~(\ref{eq:minus}) for the 
$A$=4 hypernuclei, we note that the SI $\Lambda N\leftrightarrow\Sigma N$ 
coupling potential in the LO chiral EFT $YN$ model of Ref.~\cite{polinder06} 
consists of a pseudoscalar (PS) meson exchange, dominated by OPE, plus 
two $s$-wave interaction contact terms (CT) of which the $^3S_1$ CT is 
negligible and the $^1S_0$ CT is large. In a zeroth-order single-particle 
description of the $A$=4 hypernuclei, and using Eq.~(\ref{eq:OME}), these 
$\Lambda N\leftrightarrow\Sigma N$ coupling-potential components contribute 
to the CSB separation-energy differences as follows: 
\begin{equation} 
\Delta B^{J=0}_{\Lambda} = \frac{3}{2}C_1-\frac{1}{2}C_0,\;\;\;\; 
\Delta B^{J=1}_{\Lambda} = \frac{1}{2}C_1+\frac{1}{2}C_0,  
\label{eq:CTa} 
\end{equation} 
with $C_S=C_S^{\rm CT}+C_S^{\pi}$ the sum of contributions to the 
triplet ($S$=1) and singlet ($S$=0) matrix elements from CT and from OPE. 
The ${\vec\sigma}_Y\cdot{\vec\sigma}_N$ spin dependence of $C_S^{\pi}$ 
leads in this approximation to $C_1^{\pi}=-\frac{1}{3}C_0^{\pi}$. 
Recalling that these matrix elements already incorporate isospin, both 
$C_0^{\pi}$ and $C_0^{\rm CT}$ are negative. Hence, 
\begin{equation} 
\Delta B^{J=0}_{\Lambda} \approx -\,(C_0^{\pi}+\frac{1}{2}C_0^{\rm CT}) > 0, 
\label{eq:CTb1} 
\end{equation} 
\begin{equation} 
\Delta B^{J=1}_{\Lambda} \approx +\,(\frac{1}{3}C_0^{\pi} + 
\frac{1}{2}C_0^{\rm CT}) < 0, 
\label{eq:CTb2} 
\end{equation} 
in agreement with the signs of the calculated CSB splittings. In the 
limit that $C_0^{\pi}$ is negligible with respect to $C_0^{\rm CT}$, 
Eq.~(\ref{eq:minus}) is recovered. We conclude that it is the sizable $^1S_0$ 
$\Lambda N\leftrightarrow\Sigma N$ coupling-potential CT in the LO chiral EFT 
$YN$ interaction model \cite{polinder06} that makes it possible to generate 
sufficiently large values of $\Delta B^{J=0}_{\Lambda}$ to explain the 
observed CSB splitting of the $0^+_{\rm g.s.}$ mirror levels. However, 
the opposite-sign values of $\Delta B^{J=1}_{\Lambda}$ appear too large 
with respect to the near degeneracy observed for the $1^{+}_{\rm exc.}$ 
mirror levels, even when updated values from the latest MAMI measurement 
are considered \cite{MAMI16}. 

In contrast to the ability of the LO chiral EFT $YN$ interaction model to 
generate sizable CSB g.s. splittings $\Delta B^{J=0}_{\Lambda}$ owing to 
a dominant $^1S_0$ $\Lambda N\leftrightarrow\Sigma N$ coupling-potential CT, 
the $\Lambda N\leftrightarrow\Sigma N$ coupling potential in NSC97 models 
is dominated by a $^3S_1 - {^3D_1}$ tensor component which is ineffective 
in generating a large CSB contribution when used in the right-hand side 
of Eq.~(\ref{eq:OME}). The reason is that the SI $\Lambda N$ states on 
the left-hand side, in the case of NSC97, are dominated by purely $s$-wave 
channels \cite{nogga01}. The NSC97 $^1S_0$ $\Lambda N\leftrightarrow\Sigma N$ 
coupling-potential contribution that replaces $C_0^{\pi}$ in 
Eq.~(\ref{eq:CTb1}) is too weak to generate on its own a sizable $\Delta 
B^{J=0}_{\Lambda}$. A detailed account of this item will be given elsewhere. 

It is tempting to speculate on the $A$=4 CSB separation-energy differences 
$\Delta B^J_{\Lambda}$ anticipated from applying Eq.~(\ref{eq:OME}) to the 
recently published NLO chiral EFT $YN$ interaction \cite{haidenbauer13}. 
The $\Lambda N\leftrightarrow\Sigma N$ coupling-potential contact terms differ 
considerably in NLO from those in LO, with a very large $C_1^{\rm CT}$ that 
dominates in NLO over $C_0^{\rm CT}$, and with a new $^3S_1$--$^3D_1$ CT. 
It is fair to assume that PS one- and two-meson exchange contributions in 
NLO are still dominated by OPE. Dominance of $C_1^{\rm CT}$ over all other 
allowed contributions would result in {\it negative} values of $\Delta 
B^J_{\Lambda}$, with $\Delta B^{J=0}_{\Lambda}$ 3 times as large as 
$\Delta B^{J=1}_{\Lambda}$; this would disagree with the observed positive 
value for $\Delta B^{J=0}_{\Lambda}$, see Fig.~\ref{eq:OME}, confirmed 
also by the new MAMI measurement \cite{MAMI15}. We note, furthermore, 
that the NLO version underestimates the $A$=4 hypernuclear g.s. separation 
energy, with $B_{\Lambda}^{\rm CS}\approx$1.5--1.6~MeV \cite{nogga13}, 
compared to $\approx$2.2~MeV from Fig.~\ref{fig:exp}. Three-body $YNN$ 
interaction terms introduced in higher-order versions in order to recover the 
missing g.s. attraction might provide additional source of CSB in $\Lambda$ 
hypernuclei. However, expecting that the dominant $YNN$ terms correspond 
to $\Sigma^{\ast}(1385)NN$ intermediate states \cite{petschauer15} and 
realizing that, unlike $\Sigma^0$, $\Sigma^{\ast 0}({\frac{3}{2}}^+)$ 
cannot mix with $\Lambda^0({\frac{1}{2}}^+)$ to generate CSB, these $YNN$ 
interaction terms will not produce as strong CSB as evaluated here using 
Eq.~(\ref{eq:OME}), which is based on the Dalitz--von Hippel $\Lambda^0 - 
\Sigma^0$ mixing mechanism \cite{DvH64}. It is therefore questionable whether 
the NLO version \cite{haidenbauer13} offers an advantage over the LO version 
\cite{polinder06} for $\Lambda$ hypernuclei, given also that both provide 
comparably reasonable fits to the low-energy $YN$ scattering data. 
\newline 
\noindent 
{\bf Summary and outlook.}~~In conclusion, we have presented the first CSB 
{\it ab initio} calculation in hypernuclei with chiral EFT coupled-channel 
$YN$ interactions, showing that the LO version \cite{polinder06} is capable 
of producing a {\it large} CSB $0^+_{\rm g.s.}$ splitting ${\overline{
\Delta B}}^{J=0}_{\Lambda}\sim 180\pm 130$~keV. This is consistent with a g.s. 
splitting of 270$\pm$95~keV reported by the MAMI experiment \cite{MAMI15}. 
Our NCSM calculation reproduces quantitatively and with weak cutoff dependence 
the $0^+_{\rm g.s.}$ binding energies of the $A$=4 mirror hypernuclei, 
whereas the $1^{+}_{\rm exc.}$ binding-energy calculation, which is known 
to be numerically more challenging \cite{nogga01}, displays a strong 
cutoff dependence. The calculated CSB $1^{+}_{\rm exc.}$ splitting is of 
opposite sign to that of the $0^+_{\rm g.s.}$ splitting and fairly large: 
${\overline{\Delta B}}^{J=1}_{\Lambda}\approx -200\pm 30$~keV, with a weak 
cutoff dependence. While the latest results from MAMI suggest a smaller 
negative CSB splitting of $-83\pm 94$~keV for the $1^{+}_{\rm exc.}$ 
mirror levels \cite{MAMI16}, the measurement systematic uncertainty is still 
too large to rule out the prediction of the LO version. 

In future work it would be of great interest to apply the CSB generating 
equation (\ref{eq:OME}) in {\it ab initio} calculations of the $A$=4 mirror 
hypernuclei using the recent NLO EFT version \cite{haidenbauer13}, and also 
to readjust the $\Lambda N\leftrightarrow\Sigma N$ contact terms in NLO by 
imposing the accurate CSB datum $E_{\gamma}(_{\Lambda}^4{\rm He})-E_{\gamma}
(_{\Lambda}^4{\rm H})=0.32\pm 0.02$~MeV, so it is reproduced in four-body 
calculations with as weak cutoff dependence as possible. Another natural 
follow-up would be to extend these CSB calculations in LO and NLO to $p$-shell 
hypernuclei. Recent shell model calculations \cite{gal15}, using a schematic 
$\Lambda N\leftrightarrow\Sigma N$ coupling-potential model, suggest that CSB 
splittings of g.s. mirror levels in $p$-shell hypernuclei decrease in size 
with respect to $A=4$, and perhaps even reverse sign, in rough agreement with 
old emulsion data \cite{davis05}. Such extensions of the present work pose 
a valuable theoretical challenge to the microscopic understanding of strange 
nuclear systems. 
\newline 
\noindent 
{\bf Acknowledgments}.~~We are grateful to Petr Navr\'{a}til for providing 
valuable advice and help on extensions of nuclear-physics NCSM codes, to 
Johann Haidenbauer and Andreas Nogga for providing us with the input LO EFT 
$YN$ potentials used in the present work, to Nir Barnea for stimulating 
discussions on EFT few-body applications, and to Ji\v{r}\'{i} Mare\v{s} for 
a critical reading of this manuscript. The research of D.G. was supported by 
the Grant Agency of the Czech Republic (GACR) Grant No. P203/15/04301S.


\begin{thebibliography}{99} 

\bibitem{LQCD15} S.~Borsanyi {\it et al.}, Science {\bf 347}, 1452 (2015). 

\bibitem{LQCD13} S.R.~Beane {\it et al.} (NPLQCD Collaboration), Phys. Rev. 
C {\bf 88}, 024003 (2013). 

\bibitem{entem03} D.R.~Entem and R.~Machleidt, Phys. Rev. C {\bf 68}, 
041001(R) (2003).  

\bibitem{polinder06} H.~Polinder, J.~Haidenbauer, and U.-G.~Mei{\ss}ner, 
Nucl. Phys. A {\bf 779}, 244 (2006). 

\bibitem{haidenbauer13} J.~Haidenbauer, S.~Petschauer, N.~Kaiser, 
U.-G.~Mei{\ss}ner, A.~Nogga, and W.~Weise, Nucl. Phys. A {\bf 915}, 24 (2013). 

\bibitem{davis05} D.H.~Davis, Nucl. Phys. A {\bf 754}, 3c (2005). 

\bibitem{E13} T.O.~Yamamoto {\it et al.} (J-PARC E13 Collaboration), 
Phys. Rev. Lett. {\bf 115}, 222501 (2015). 

\bibitem{MAMI15} A.~Esser {\it et al.} (A1 Collaboration), Phys. Rev. Lett. 
{\bf 114}, 232501 (2015). 

\bibitem{miller06} G.A.~Miller, A.K.~Opper, and E.J.~Stephenson, Annu. Rev. 
Nucl. Part. Sci. {\bf 56}, 253 (2006). 

\bibitem{DvH64} R.H.~Dalitz and F.~von Hippel, Phys. Lett. {\bf 10}, 153 
(1964).  

\bibitem{gal15} A.~Gal, Phys. Lett. B {\bf 744}, 352 (2015). 

\bibitem{gal15a} see Comment by A.~Gal, Phys. Rev. D {\bf 92}, 018501 (2015) 
and Reply by R.~Horsley {\it et al.} (QCDSF-UKQCD Collaboration), Phys. Rev. 
D {\bf 92}, 018502 (2015), where the incompleteness of existing LQCD 
calculations of $\langle\Sigma^0|\delta M|\Lambda\rangle$ is discussed. 

\bibitem{NSC97} Th.A.~Rijken, V.G.J.~Stoks, and Y.~Yamamoto, Phys. Rev. C 
{\bf 59}, 21 (1999). 

\bibitem{nogga01} A.~Nogga, Ph.D. Dissertation submitted to the Ruhr-
\newline 
\noindent 
Universit\"{a}t, Bochum (2001), http://www-brs.ub.ruhr-
\newline 
\noindent 
uni-bochum.de/netahtml/HSS/Diss/NoggaAndreas/. 

\bibitem{nogga02} A.~Nogga, H.~Kamada, and W.~Gl\"{o}ckle, Phys. Rev. Lett. 
{\bf 88}, 172501 (2002). 

\bibitem{haidenbauer07} J.~Haidenbauer, U.-G.~Mei{\ss}ner, A.~Nogga, and 
H.~Polinder, in {\it Topics in Strangeness Nuclear Physics}, Lecture Notes 
in Physics {\bf 724}, edited by P.~Byd\v{z}ovsk\'{y}, J.~Mare\v{s}, and 
A. Gal (Springer, New York, 2007), pp.~113-140. 

\bibitem{nogga13} A.~Nogga, Nucl. Phys. A {\bf 914}, 140 (2013) and 
references to earlier works cited therein. 

\bibitem{nogga14} A.~Nogga, Few-Body Syst. {\bf 55}, 757 (2014). 

\bibitem{ESC10} Th.A.~Rijken, M.M.~Nagels, and Y.~Yamamoto, 
Prog. Theor. Phys. Suppl. {\bf 185}, 14 (2010). 

\bibitem{kyoto07} Y.~Fujiwara, Y.~Suzuki, and C.~Nakamoto, 
Prog. Part. Nucl. Phys. {\bf 58}, 439 (2007). 

\bibitem{coon99} S.A.~Coon, H.K.~Han, J.~Carlson, and B.F.~Gibson, 
in {\it Meson and Light Nuclei '98}, edited by J.~Adam, P.~Byd\v{z}ovsk\'{y}, 
J.~Dobe\v{s}, R.~Mach, and J.~Mare\v{s} (WS, Singapore, 1999), pp.~407-413, 
arXiv:nucl-th/9903034. 

\bibitem{navratil00} P.~Navr\'{a}til, G.P.~Kamuntavi\v{c}ius, 
and B.R.~Barrett, Phys. Rev. C {\bf 61}, 044001 (2000). 

\bibitem{gazda14} D.~Gazda, J.~Mare\v{s}, P.~Navr\'{a}til, R.~Roth, and 
R.~Wirth, Few-Body Syst. {\bf 55}, 857 (2014). 

\bibitem{wirth14} R.~Wirth, D.~Gazda, P.~Navr\'{a}til, A.~Calci, 
J.~Langhammer, and R.~Roth, Phys. Rev. Lett. {\bf 113}, 192502 (2014). 
Note that with the same Jacobi-coordinate NCSM methodology and the same 
$NN$ and $YN$ interactions used in the present paper, these authors found 
$_{\Lambda}^3$H to be particle-stable, with $\Lambda$ separation energy 
$B_{\Lambda}$=110$\pm$10~keV for cutoff 600 MeV, consistent with experiment 
\cite{davis05} and with Faddeev calculations reported by Haidenbauer 
{\it et al.} \cite{haidenbauer07}. CSB from $\Lambda - \Sigma^0$ mixing has 
no effect in $_{\Lambda}^3$H calculations. 

\bibitem{navratil07} P.~Navr\'{a}til, Few-Body Syst. {\bf 41}, 117 (2007).

\bibitem{GL88} B.F.~Gibson and D.R.~Lehman, Phys. Rev. C {\bf 37}, 679 (1988). 

\bibitem{akaishi00} Y.~Akaishi, T.~Harada, S.~Shinmura, and K.S.~Myint, 
Phys. Rev. Lett. {\bf 84}, 3539 (2000). 

\bibitem{hiyama02} E.~Hiyama, M.~Kamimura, T.~Motoba, T.~Yamada, 
and Y.~Yamamoto, Phys. Rev. C {\bf 65}, 011301(R) (2002). 

\bibitem{akaishi02} H.~Nemura, Y.~Akaishi, and Y.~Suzuki, Phys. Rev. Lett. 
{\bf 89}, 142504 (2002). 

\bibitem{MAMI16} F.~Schulz {\it et al.} (A1 Collaboration), Nucl. Phys. A 
{\bf 954}, 149 (2016). Adopting their value of $B_{\Lambda}(_{\Lambda}^4{
\rm H})=2.157\pm 0.006({\rm stat.})\pm 0.077({\rm syst.})$~MeV, the observed 
CSB splittings would change to $233\pm 92$~keV for $0^+_{\rm g.s.}$ and to 
$-83\pm 94$~keV for $1^{+}_{\rm exc.}$. We thank Dr. Patrick Achenbach for 
briefing us of these values prior to publication. 

\bibitem{petschauer15} S.~Petschauer, N.~Kaiser, J.~Haidenbauer, 
U.-G.~Mei{\ss}ner, and W.~Weise, Phys. Rev. C {\bf 93}, 014001 (2016), 
and references cited therein to earlier work. 


\end{thebibliography}
\end{document}